# Degeneracy in the genetic code and its symmetries by base substitutions.

Jean-Luc Jestin

*Unité de Chimie Organique, Département de Biologie Structurale et Chimie, Institut Pasteur*

*28 rue du Dr. Roux, 75724 Paris cedex 15, France*

**Abstract**

**Why is the genetic code the way it is? The most successful theory states that the codon assignments minimise the effects of errors arising in primordial living systems (1). Here a transversion is reported that leaves invariant degeneracy in the genetic code. A model was established to address the question: How should sets of synonymous codons be chosen to minimise the deleterious effects of non synonymous substitutions? It is shown that the arithmetic theory of quadratic forms provides the framework for degeneracy in the genetic code.**

Primordial living systems require the cooperation between nucleic acids and proteins as shown by RNA-directed protein synthesis and by polymerase-catalysed nucleic acid amplification (2-4). Reducing the effects of biochemical errors in these primordial systems has often been suggested as the major force shaping the genetic code (1,5-8). Symmetries of the degeneracy pattern by base substitutions are intrinsic properties of the genetic code (9). They suggested the use of group theory as a mathematical formalism (10). These symmetries are shown here to derive algebraically from the calculation of non synonymous base substitutions for sets of synonymous codons.

Let us consider the general case of a degenerate code of n codons that are k-bases long, $(b_1b_2...b_k)$. The degeneracy pattern is such that there are $\omega$ sets of synonymous codons. The average number S of non-synonymous substitutions over the entire coding sequence of the replicating system can be written as a function of the number of non-synonymous substitutions for each set **a** of synonymous codons:

$$\sum_{a=1}^{\omega} S_a = S \qquad [1]$$

Let us formulate the number of non-synonymous substitutions $S_a$ for a set of synonymous codons **a** as a function of the number of substitutions of any codon i into any codon j. It is the sum over all synonymous codons of the number of mutations to any codon that does not belong to the set of synonymous codons:

$$\tilde{\mathbf{x}}_a \, M \, \mathbf{y}_a = S_a \qquad [2]$$

M is the nxn matrix with coefficients $m_{ij}$, which are the number of substitutions of codon i into codon j ($m_{ij} = 0$ for $i = j$). This general formalism takes into account the fact that the fidelity of nucleic acid polymerases is highly dependent on sequence context (11-15). A related formalism has been used to estimate mutational 'deterioration' for any codon substitution (16). $\mathbf{x}_a$ is the vector $(x_1, x_2..., x_n)$ with $x_i = 1$ if codon i belongs to the set **a** of synonymous codons or otherwise $x_i = 0$; the symbol ~ on a vector indicates its transpose, **1** is the n-tuple (1, 1,…, 1), $\mathbf{y}_a$ the vector $(y_1, y_2, \ldots, y_n)$ and $\mathbf{y_a} = \mathbf{1} - \mathbf{x_a}$. tRNA-mis-aminoacylation which alters the number of non-synonymous substitutions, can be further taken into account as factors correcting $\mathbf{x_a}$ and $\mathbf{y_a}$. For each codon i belonging to the set $\mathbf{a}$ of synonymous codons, mis-translation by this mechanism is equivalent to setting $x_i = \alpha_i$ with $\alpha_i$ taken as a rational close to 1, higher than 1 and varying with codon i belonging to the set $\mathbf{a}$ of synonymous codons; otherwise $\alpha_i = 0$ for codon i not belonging to the set $\mathbf{a}$ of synonymous codons. Furthermore, the substitution of codon i of the set $\mathbf{a}$ into the non-synonymous codon j combined with an appropriate tRNA-mis-aminoacylation is equivalent to a correct translation of codon i and therefore not counted as a non-synonymous substitution: $\beta_j$ is introduced as a rational close to 1, smaller than 1 and varying with codon j such that $y_j = \beta_j$ or otherwise $y_i = 0$ if codon i belongs to the set $\mathbf{a}$. The left part of equation [2] is a quadratic form, whose 2nx2n matrix is symmetric (17). The quadratic form is considered over the set $\mathbf{Q}$ of rational numbers.

For two quadratic forms to be equivalent over $\mathbf{Q}$, it is necessary and sufficient that they are equivalent over each p-adic field $\mathbf{Q_p}$ and over $\mathbf{R}=\mathbf{Q_\infty}$ (17-19). The p-adic distance characterising the field $\mathbf{Q_p}$ is an ultrametric measure. For example, two integers a and b are near p-adically if their decomposition into prime numbers have in common a high power of p; their distance can be defined as $1/p^\alpha$, where $\alpha$ is highest common power of p. It satisfies the relation: $\text{norm}(a + b) \leq \sup[\text{norm}(a), \text{norm}(b)]$. The link between such measures and phylogenetic trees was established (20). The need to define nearness relations in molecular evolution has been pointed out (21). The number of substitutions from codon i to codon j is proportional to the product for all codon bases of the probabilities for each base to mutate or not in a given sequence context. Two such numbers can be defined as near if they have a probability as a common factor.

So as to be independent of the vector basis chosen, including the order of the n codons, the class of equivalent quadratic forms associated to the set $\mathbf{a}$ of synonymous codons is further considered. A class of quadratic forms over $\mathbf{Q_p}$ is defined by its invariants, which are respectively the rank, the determinant $\mathbf{d}$ which is part of the quotient group $\mathbf{Q^*_p}/\mathbf{Q^*_p}^2$ and the Hasse invariant $\varepsilon$ which equals $\pm 1$ (17-19).

The degeneracy pattern of the genetic code can then be mapped to the invariants of the set of classes of quadratic forms of dimension $\mathbf{n} \geq 3$ over $\mathbf{Q_p}$ (Figs. 1 and 2). The invariants characterise the sets of synonymous codons. The set of 64 codons can be divided into a set of 32 codons noted group (IV) whose third base does not have to be specified so as to define an amino acid. For the other 32 codons, noted group (II), three bases have to be specified so as to define unambiguously an amino acid or a stop signal. Rumer observed that there exists a unique symmetry exchanging group (IV) into group (II) which substitutes G into T, T into G, C into A or A into C and which applies to all codon bases (9). Another unique symmetry described here exchanges each group into itself by substitution of G into C, C into G, A into T or T into A applied to the first codon base (Fig. 2.1). It is also clear for the vertebrate mitochondrial genetic code that a transition applied to the third codon base is a synonymous mutation (Fig. 2.2). Given a codon assigned to a group, the three symmetries define a set of three codons whose group assignments are then predicted. The three symmetries are associated to all three types of base substitutions, transitions and transversions. For $p \neq 2$, there are 8 classes of quadratic forms over $\mathbf{Q}_p$ (cf. group (IV)) and for $p=2$ there are 16 classes (cf. group (II)). The Hasse invariant $\varepsilon$ which equals $\pm 1$, can be associated to the unique symmetry exchanging each group into itself. The determinant has 4 representatives for $p \neq 2$ (cf. group (IV)) and 8 representatives for $p=2$ (cf. group (II)). The determinant belongs to the quotient group $\mathbf{Q}^*_p/\mathbf{Q}^{*2}_p$ which can be considered for $p \neq 2$ as a two dimensional vector space over the field $\mathbf{F}_2$ of two elements (cf. group (IV) for which two bases have to be defined) and for $p=2$ as a three dimensional vector space over $\mathbf{F}_2$ (cf. group (II) for which three bases have to be defined).

In this model, the four bases are associated to the four elements of $\mathbf{F}_2 \times \mathbf{F}_2$ for both conditions $p=2$ and $p \neq 2$. This model confirms that three base codons are sufficient. It puts also an upper limit to the number $\omega$ of synonymous codon sets present in the genetic code of 24 = 16 + 8. Twenty amino acids belong to the standard set of amino acids, as one set of synonymous codons is attributed to the stop signal and as the three amino acids encoded by six codons in the standard genetic code Leu, Arg and Ser are considered each as two sets of synonymous codons, one belonging to group (IV) and one belonging to group (II). The number of amino acids coded within the genetic code varies among the various known genetic codes: their diversity has been described by evolutionary models (22, 23) and can be created experimentally (24). It is often associated to start signals and to stop signals whose codon assignment was observed to optimise the tolerance of polymerase-induced frameshift mutations (25).

Minimisation of non-synonymous substitutions may have had a central role for the assignment of amino acids to sets of synonymous codons as defined above.

Links between coding and the genetic code have been established (26, 27). The theorem on the classification of quadratic forms over the field of rational p-adic numbers can be considered as an extension of coding theory to degenerate codes whose codons' error rates are sequence dependent. The theorem provides the algebraic framework to define sets of synonymous codons together with the relations between these sets, as found in the quasi-universal degeneracy pattern of the genetic code.

**References.**

**Acknowledgements.**

The author is greatly indebted to G. Christol, A. Kempf and F. Guichard and thanks F. Courtes, Y. Benoist, S. Wain-Hobson, A. Ullmann and V. Shcherbak for discussions and for criticisms. Library facilities at the Ecole Normale Supérieure and a grant from the Ministère de la Recherche (MENRT, ACI blanche) are acknowledged.

**Figure legends.**

**Figure 1.**

p is a prime number, $Q_p$ the field of p-adic rational numbers, ε the Hasse symbol, d the determinant, $F_2$ the field of two elements.

**Figure 2.1.**

Group (IV) contains all codons for which the third codon base (N) does not have to be defined to specify an amino acid. Group(II) contains the other 32 codons. Curved arrows indicate substitutions applied to a single base and the straight arrow, a substitution applied to the three codon bases. For example, (GC/AT) indicates the substitution of G into C, C into G, A into T or T into A. N = A,T,G or C; Y = T or C; R = A or G.

**Figure 2.2.**

A hyphen indicates a stop signal.

## Figure 1. The classification of quadratic forms of dimension n≥3 over $Q_p$

For p≠2, there are **8** classes of quadratic forms. Their invariants are:

$\varepsilon = \pm 1$

**d** in $Q^*_p/Q_p^2$ which has **4** representatives
and which can be considered as a **2**-dimensional vector space over $F_2$

For p=2, there are **16** classes of quadratic forms. Their invariants are:

$\varepsilon = \pm 1$

**d** in $Q^*_p/Q_p^2$ which has **8** representatives
and which can be considered as a **3**-dimensional vector space over $F_2$

## Figure 2. Two representations of the genetic code's degeneracy pattern

### 2.1. A representation of the transformations conserving or exchanging the degeneracy pattern

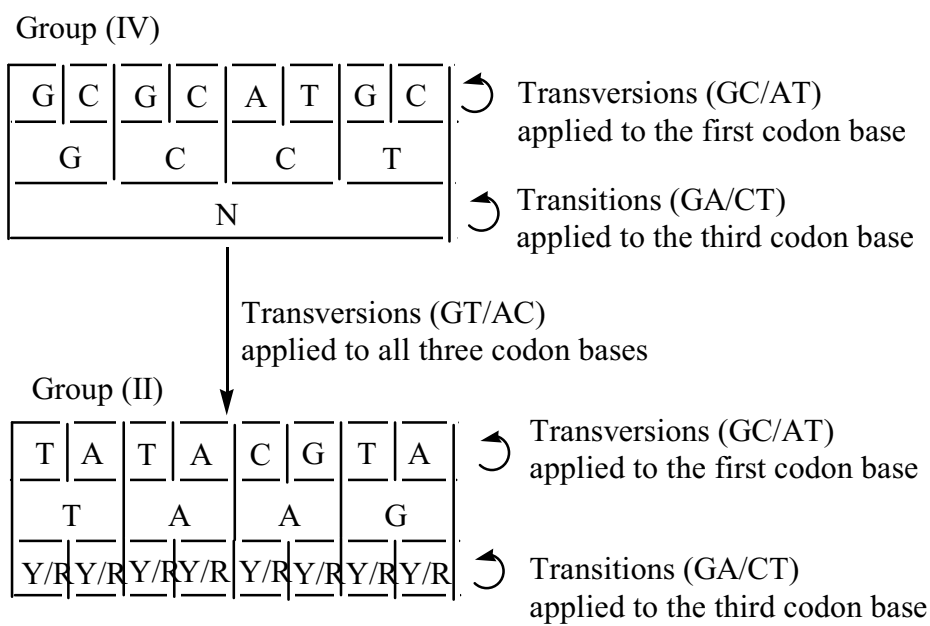

Group (IV)

Transversions (GC/AT) applied to the first codon base

Transitions (GA/CT) applied to the third codon base

Transversions (GT/AC) applied to all three codon bases

Group (II)

Transversions (GC/AT) applied to the first codon base

Transitions (GA/CT) applied to the third codon base

### 2.2. A representation of the vertebrate mitochondrial genetic code

```
Gly        Arg        Ala        Pro        Thr        Ser        Val        Leu

GGG    /   CGG        GCG    /   CCG        ACG    /   TCG        GTG    /   CTG
GGA        CGA        GCA        CCA        ACA        TCA        GTA        CTA
GGC        CGC        GCC        CCC        ACC        TCC        GTC        CTC
GGT        CGT        GCT        CCT        ACT        TCT        GTT        CTT

Phe\Leu    Ile\Met    Tyr\-      Asn\Lys    His\Gln    Asp\Glu    Cys\Trp    Ser\-

TTT\TTG /  ATT\ATG    TAT\TAG /  AAT\AAG    CAT\CAG /  GAT\GAG    TGT\TGG    /  AGT\AGG
TTC TTA    ATC ATA    TAC TAA    AAC AAA    CAC CAA    GAC GAA    TGC TGA       AGC AGA
```